\documentclass[twocolumn,showpacs,preprintnumbers,amsmathamssymb]{revtex4}
\usepackage{graphicx}
\usepackage{epsfig}
\preprint{Superlattices}
\begin{document}

\title{Optical properties of (SrMnO$_3$)$_n$/(LaMnO$_3$)$_{2n}$ superlattices: an insulator-to-metal transition observed in the absence of disorder}
\author{A. Perucchi$^1$, L. Baldassarre$^1$, A. Nucara$^2$, P. Calvani$^2$, C. Adamo$^3$, D.G. Schlom$^3$, P. Orgiani$^4$, L. Maritato$^4$, and S. Lupi$^5$}
\affiliation{$^1$Sincrotrone Trieste, Area Science Park, I-34012 
Trieste, Italy}
\affiliation{$^2$CNR-SPIN and Dipartimento di Fisica, Universit\`a 
di Roma La Sapienza, Piazzale Aldo Moro 2, I-00185 Rome, Italy}
\affiliation{$^3$ Department of Materials Science and Engineering, Cornell University, Ithaca, New York 14853-1501, USA}
\affiliation{$^4$ CNR-SPIN and Dipartimento di Matematica ed Informatica, Universit\`a  di Salerno, Baronissi, I-84081 Salerno, Italy}
\affiliation{$^5$CNR-IOM and Dipartimento di Fisica, Universit\`a 
di Roma La Sapienza, Piazzale Aldo Moro 2, I-00185 Rome, Italy}

\date{\today}

\begin{abstract}
We measure the optical conductivity $\sigma_1 (\omega)$ of (SrMnO$_3$)$_n$/(LaMnO$_3$)$_{2n}$ superlattices (SL) for $n=1,3,5$, and 8 and  $10 < T < 400$ K. Data show a $T$-dependent insulator to metal transition (IMT) for $n \leq 3$, driven by the softening of a polaronic mid-infrared band.  At $n$ = 5 that softening is incomplete, while at the largest-period $n=8$ compound the MIR band is independent of $T$ and the SL remains insulating. One can thus first observe the IMT in a manganite system in the absence of the disorder due to chemical doping. Unsuccessful reconstruction of the SL optical properties from those of the original bulk materials suggests that (SrMnO$_3$)$_n$/(LaMnO$_3$)$_{2n}$ heterostructures give rise to a novel electronic state.

\end{abstract}

\maketitle

Since decades, manganites attract the greatest attention of the condensed matter community \cite{tokura99,tokura06}. This is due either to the Colossal Magnetoresistance (CMR) which makes them appealing for the applications, and to the  rich  doping-temperature phase diagram which is of great interest for basic research. In particular, several studies have been devoted to the Insulator-to-Metal Transition (IMT) coupled to ferromagnetic ordering at hole-doping levels $\approx 1/3$. The IMT is understood through the double-exchange (DE) mechanism \cite{zener51}, once the localization tendency due to polaron formation has been taken into account \cite{millis95}. Today it is widely recognized that quenched disorder weakens long range order and causes ferromagnetism to break up into clusters, whose sudden alignment in the presence of a magnetic field is an essential ingredient of CMR  \cite{tokura06,dagotto05}.

Recent progress in the growth of atomic-scale multilayers opens exciting opportunities in the design of  materials with novel properties. The so-called "electronic reconstruction" effect produces a new 2-D metallic state at the interface between a band insulator as SrTiO$_3$ and a Mott insulator like  LaTiO$_3$  \cite{ohtomo02,okamoto04,seo07}. Manganite superlattices with alternating layers of insulating anti-ferromagnets SrMnO$_3$ (SMO) and LaMnO$_3$ (LMO) have been recently studied as well \cite{smadici07,bhattacharya08,adamo09,aruta09,may09}. Thanks to electronic reconstruction, both metallicity and ferromagnetism can be induced in these nano-structured superlattices (SLs) \cite{lin06,lin08,dong08}. 

The (SrMnO$_3$)$_n$/(LaMnO$_3$)$_{2n}$ SL system corresponds to an effective hole doping 1/3.  When $n<3$, the spacing between interfaces is so small, that the 3-D charge distribution throughout the film is believed to be uniform \cite{adamo09,lin08}. For $n=1$  an IMT below a temperature $T_{IMT}$ comparable to that of the corresponding La$_{2/3}$Sr$_{1/3}$MnO$_3$ takes place in the absence of random disorder \cite{adamo09}. 
For sufficiently low temperatures, dc measurements show that  the film made up of two insulators becomes metallic  for small $n$ \cite{adamo09}. Nevertheless, the way the IMT is approached for decreasing $n$, the excitations involved,  and the SL low-energy electrodynamics are still to be investigated. One may also ask whether the dc resistivity reflects the carrier dynamics at the interfaces or that in between,  and whether the optical properties resemble or not those of the bulk materials. We address these problems by means of infrared spectroscopy, a tool which has been successfully used to probe the low-energy electrodynamics \cite{dressel} and the mechanism of the IMT \cite{lupi09} in a number of oxides. The results on (SrMnO$_3$)$_n$/(LaMnO$_3$)$_{2n}$ superlattices with $n=1,3,5,8$ are presented here, in a wide range of frequency and temperature. 

\begin{figure}
{\hbox{\psfig{figure=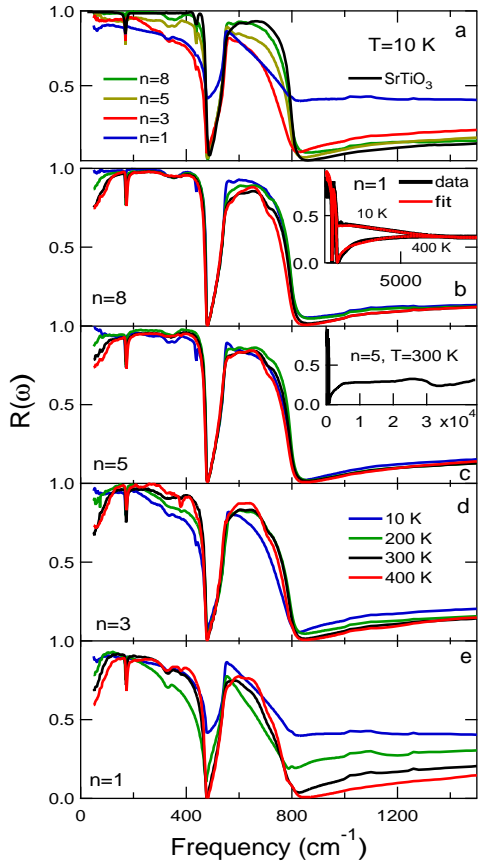,width=7cm}}} 
\caption{(Color online) Reflectivity measurements of the (SMO)n/(LMO)2n superlattices on STO. (a)  $R(\omega)$ at 10 K of the STO bare substrate, together with that of  the $n=1$, 3 , 5 and 8 SLs on STO. (b) $T$-dependent $R(\omega)$ of the $n=8$ film on STO. (c), (d) and (e) Same for $n=5$, $n=3$ and $n=1$ respectively. The legend shown in panel (d), applies to panels (b), (c), and (e), as well. Inset b: Data and fits at 10 and 400 K for $n=1$. Inset c: $R(\omega)$ for $n=5$ up to 45000 cm$^{-1}$.}
 \label{Reflectivity}
\end{figure}

The four  films of (SrMnO$_3$)$_n$/(LaMnO$_3$)$_{2n}$, all 20 nm thick, have been grown by molecular-beam epitaxy on SrTiO$_3$ (STO) substrates, following a procedure described elsewhere \cite{adamo09}. We have measured their reflectivity ($R(\omega)$) from 50 to 45000 cm$^{-1}$ at nearly normal incidence, between 10 and 400 K, employing either Au or Al reference mirrors. The results at 10 K are reported and compared with each other in Fig. \ref{Reflectivity}-a, together with the $R(\omega)$ of a bare STO substrate, etched and annealed under the same conditions used for the preparation of the SLs. Its contribution obviously dominates the shape of $R(\omega)$  in all films, and particularly for a large-period superlattice such as  $n=8$. However, with decreasing $n$ the reflectivity decreases appreciably below 800 cm$^{-1}$ and, at $n=1$, the shape of $R(\omega)$ has substantially changed   indicating that such SL does effectively screen the substrate. Indeed, a similar $R(\omega)$  is reported in Ref.   \onlinecite{haghiri08} for a  La$_{2/3}$Sr$_{1/3}$MnO$_3$ film grown on STO, at low temperature.
Panels from b) to e) in Fig. \ref{Reflectivity} show the $T$-dependence of  $R(\omega)$ in the individual SLs. The effect observed below 150 cm$^{-1}$ in all samples is not related to the SL physics, being due to the STO substrate approaching the ferroelectric phase \cite{dore97}. At higher energies, for large $n$ (Fig. \ref{Reflectivity}-b and -c) the $T$-dependence is poor, and similar to that of the STO substrate. A more pronounced evolution is observed for $n$ = 3  in Fig. \ref{Reflectivity}-d and especially  for  $n=1$, for which  $R(\omega)$ drastically changes over the whole range (Fig. \ref{Reflectivity}-e). Finally, all curves converge at about 8000 cm$^{-1}$ as shown in the inset. 

\begin{figure}
{\hbox{\psfig{figure=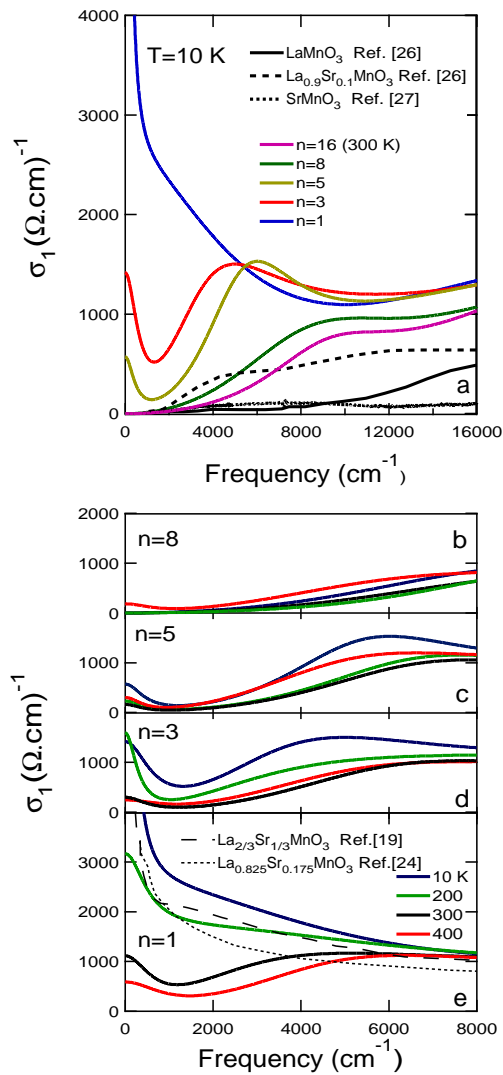,width=7cm}}} 
\caption{(Color online) $\sigma_1(\omega)$ as extracted from a Drude-Lorentz fit to the reflectivity. a) $\sigma_1(\omega)$ at $T=10$ K, for the $n=1,3,5,8$ compounds, showing the Mott transition induced by the proximity between the layers. $\sigma_1(\omega)$ at $T=300$ K, for $n=16$ is reported as well. Data on single crystals of LaMnO$_3$ and La$_{0.9}$Sr$_{0.1}$MnO$_3$ (from Ref. \cite{okimoto97}) and on SrMnO$_3$ (from Ref. \cite{nucara}) at 10 K are also shown for comparison. b), c), d) and e) display  $\sigma_1(\omega)$ at different $T$ for $n=8,5,3$ and 1 respectively, in comparison with the low-$T$ conductivity of cleaved La$_{0.825}$Sr$_{0.175}$MnO$_3$ single crystals \cite{takenaka99} and La$_{2/3}$Sr$_{1/3}$MnO$_3$ films \cite{haghiri08} at low temperature.
\label{sigma}}
\end{figure}

To extract the optical conductivity from our data we have performed Drude-Lorentz (DL) fits to $R(\omega)$, by taking into account the contributions from both the film and the substrate, as well as the (incoherent) internal reflections between them, by using standard formulae for thin films  \cite{dressel}. The model treats the film as an effective medium, neglecting the internal reflections from the LMO/SMO interfaces, whose depths and spacings are much smaller than any wavelength here employed.  
 Best DL fits need just four components: one Drude term, one mid-infrared (MIR) band, and two higher energy oscillators (the lowest being centered at 2 eV) which model the electronic interband transitions. 
This choice yields the same results as a Kramers-Kronig constrained analysis \cite{haghiri08,kuzmenko05}.  
The resulting real part of the optical conductivity $\sigma_1(\omega)$ is displayed in Fig. \ref{sigma} for different SL periods and different temperatures. 

\begin{figure}
{\hbox{\psfig{figure=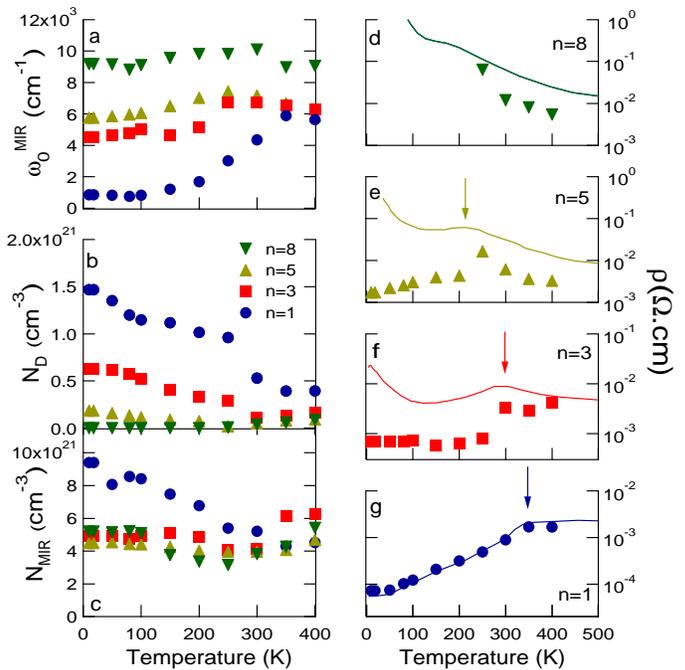,width=9cm}}} 
\caption{(Color online) Left, top: Peak frequency of the mid-infrared oscillator used in the Drude-Lorentz fit. Middle and bottom: $N_{D}$ and $N_{MIR}$  vs $T$, respectively, estimated from Eq. \ref{Ndrude} Right: the resistivity of the $n=1,3,5,8$ SLs   from Ref. \cite{adamo09} (solid lines) is compared with estimates of $\rho (T)$ from Drude-Lorentz fits to the present optical data (full symbols). The arrows mark the T$_{IMT}$ in $n=1,3$, and 5 \cite{adamo09}.} 
\label{parameters}
\end{figure}

Panel a in Fig. \ref{sigma} displays $\sigma_1(\omega)$ at 10 K, for all the SLs under consideration. The $n=8$ compound has a clear insulating gap, associated to a mid-infrared absorption band at about 1 eV, and no Drude component. For decreasing $n$ (\textit{i.e.}, decreasing SL period), a sizable Drude term appears, while the absorption edge softens appreciably. At $n=1$ we are left with a sharp Drude peak superimposed to a low-frequency harmonic oscillator centered below 1000 cm$^{-1}$, thus highlighting that a simple Drude term
does not properly describe the charge dynamics in the metallic state \cite{takenaka02}.
The presence of a broad MIR band is frequently observed in manganites and other oxides, and is a signature for charge-carriers mass renormalization.
In manganites, the source of mass enhancement is usually identified with polaron formation \cite{millis95,calvani01}.
For $n=1$, the shape of the optical conductivity is strikingly similar to that measured on  La$_{0.825}$Sr$_{0.175}$MnO$_3$ cleaved single crystals \cite{takenaka99} and La$_{2/3}$Sr$_{1/3}$MnO$_3$ films \cite{haghiri08}, as shown in Fig. \ref{sigma}-e. Therein, as $T$ increases, the Drude term weakens, while the mid-infrared band  starts to harden above $\sim 100$ K (see  Fig. \ref{parameters}), until a minimum in $\sigma_1(\omega)$ forms at about 1500 cm$^{-1}$ between the two terms.  An isosbestic point is observed at about 1 eV ($\sim 8000$ cm$^{-1}$).

We report in the left column of Fig. \ref{parameters} the temperature dependence of the peak frequency $\omega_0^{MIR}$ of the MIR band, as well as the spectral weight $N_D$ and $N_{MIR}$ of  the Drude term and the MIR band, respectively, which are defined by 

\begin{equation}
N_{D,MIR}=\frac{m_e}{4\pi e^2}\omega_{D,MIR}^2/V 
\label{Ndrude}
\end{equation}

\noindent
Here $V$ is the sample  volume, $m_e$ is the  bare electron mass, $\omega_{D}$ is the Drude plasma frequency and $\omega_{MIR}$ the oscillator strength of the MIR band.

The right column of Fig. \ref{parameters} shows instead the resistivity of the films, as measured directly in Ref. \onlinecite{adamo09} (lines) and as obtained from extrapolations to $\omega =0$ of the present conductivity data (symbols).  At high $n$ the optical values are systematically lower than the dc measurements. This discrepancy may be due to disorder-induced localization effects at sub-THz frequencies, which cannot be probed by the optical measurement \cite{dressel}. However one should also take into account that while optics probes in-plane excitations only, the dc measurement may include paths perpendicular to the layers with higher resistivity, in series to those parallel to them \cite{dong08}. Indeed, at  $n\geq 3$ the charge distribution is not believed to be uniform, since the Thomas-Fermi length-scale for charge leakage is estimated to fall between 1 and 3 unit cells \cite{lin06}.  For long SL periods, as $T$ lowers, $\rho(T)$ first displays a DE-driven IMT at the $T_{IMT}$ indicated by the arrows, and then an insulating-like variable-range-hopping \cite{adamo09,mott90} (panels d, e, f). This low-$T$ upturn is not observed in the $\rho_{opt}(T)$ extracted from the reflectivity at normal incidence, which probes the in-plane conductivity only, averaged between the various ($l+2n$) layers ($\sigma=\Sigma_{i=1}^{l}\sigma_i/l$) \cite{lin08}. For this reason, optical measurements are less affected by the more insulating out-of-plane behavior. Nevertheless, both for $n=3$ and 5, the $T_{IMT}$ values obtained by the two techniques in Fig. \ref{parameters} are consistent. For the $n=8$ SL, below 250 K the optical spectra cannot provide any more a reliable estimate of $\sigma_{dc}$.

We start our discussion  of Fig. \ref{parameters} from the largest-period $n=8$ compound. In panel a), its $\omega_0^{MIR}$  is large (about 10000 cm$^{-1}$) and independent of $T$ within errors. One may notice that neither LMO \cite{okimoto97} nor SMO \cite{nucara} alone (see Fig. \ref{sigma}-a) display electronic bands below 2 eV (16.000 cm$^{-1}$). Nevertheless, manganites at low-doping levels as La$_{0.9}$Sr$_{0.1}$MnO$_3$ exhibit a substantial absorption below 2 eV, while remaining insulating \cite{okimoto97}. Therefore,  the MIR band in the $n=8$  superlattice should result from an effective charge doping introduced by the interface, and leaked throughout the overall system, as suggested in Fig. 3 of Ref. \onlinecite{adamo09}. A similar behavior has also been observed in the $n$ = 16 SL, reported in Fig. \ref{sigma}-a at 300 K only.
In Fig. \ref{sigma}-b, for $n$ = 8 a tiny Drude term is observed above 200 K only, accounting for a carrier density on the order of 10$^{19}$ cm$^{-3}$. For decreasing $T$ the material becomes more insulating, in agreement with the semiconducting-like behavior of the resistivity in panel d). As for large $n$ the number of interfaces is rather diluted within the sample thickness, one cannot  exclude the presence of a low-$T$ metallic state in the $n=8$ SL. Nevertheless,  we can safely set an upper bound to the sheet carrier density $N_{D}^{sheet}$ of this thin layer at  5x10$^{12}$ cm$^{-2}$. 

In the opposite limit of short period ($n=1$), Fig. \ref{parameters} clearly shows the mechanism of its transition to the metallic state. In panel a), around room temperature $\omega_0^{MIR}$ starts a spectacular decrease from 5000 to about 1000 cm$^{-1}$ (at 100 K). Meanwhile its intensity increases (panel c), a strong Drude term is built up (b) and the resistivity exhibits a metallic behavior (Fig. \ref{parameters}-g).  The increase of Drude and MIR spectral weight must be at the expenses of higher energy bands, but we cannot verify this assumption for the difficulty in taking reliable data vs. temperature at high energy on such thin films. One can just say that the sum rule on the optical conductivity may not be fulfilled in these SLs unless the cutoff frequency is much larger than 10000  cm$^{-1}$.
Basing on the above considerations and on the close resemblance between the $\sigma_1(\omega)$ of the SL and of the corresponding alloy in Fig. \ref{sigma}-b,  one can assume a uniform charge density distribution perpendicularly to the layers. This carrier density can be calculated by adding to the Drude spectral weight that of the polaronic charges in the MIR band. At $n$ = 1 and low $T$  one obtains $N=N_{D}+N_{MIR}\sim5$x$10^{21}$ cm$^{-3}$, which roughly corresponds to 1/3 of the Mn ion density of the film, after using the proper band-mass values $m_b\sim0.5m_e$ \cite{mannella05}. 
We can also estimate the mass renormalization factor as  $m^*/m_{b}=\frac{N_{Drude}+N_{MIR}}{N_{Drude}}$ \cite{vanmechelen08,devreese09}. This yields $m^*/m_{b}\sim 7$, in fair agreement with previous reports \cite{haghiri08,takenaka02}. If instead of a Drude+MIR model one assumes an extended Drude framework, one finds $m^*(\omega\rightarrow 0)/m_{b}\sim 6.5$, thus demonstrating the generality of the present results.

For intermediate $n$ ($n=3$, 5 in Fig. \ref{parameters}), a weak Drude term comes out at low $T$. This is likely  due to the presence of metallic-like layers (of mixed valence) close to the interfaces, sandwiched between the insulators (layers made up of integer or almost integer Mn$^{3+}$ and Mn$^{4+}$ ions).  A proper decomposition of the optical conductivity implies a detailed knowledge of the layer-dependent charge density distribution. Since this quantity is not available we can make the simplifying assumption that the metallic layers behave as those of the $n=1$ compound. Under this hypothesis we can provide a rough estimate of the metallic fraction, as the ratio of the Drude carrier density over that of the homogeneous $n=1$ compound, thus yielding 40\% and 10\% for $n=3$ and 5 respectively. Note however, that a crude reconstruction of the optical spectra of $n=3$ and 5, as linear combinations of the optical conductivities of $n=1$, bulk LMO,  and bulk SMO does not fit the data of Fig \ref{sigma}. This makes the (SMO)n/(LMO)2n superlattice to appear as a novel member, with its own properties, of the manganite family.  

In conclusion, we have presented here the first optical measurements on (SMO)n/(LMO)2n superlattices, aimed at observing the mechanism of an IMT driven by carrier density modulation and temperature in the absence of disorder. At the largest lattice period ($n=8$)  the spectra indicate an insulating or poorly metallic behavior at low temperatures, with the presence of a MIR band around 10000 cm$^{-1}$,  which is not observed in bulk SMO or LMO and  may be attributed to strongly localized charges (small polarons).  The same band is found at lower energies both at $n$ = 5 and $n$ = 3, where its slight softening at low $T$ is accompanied by the building of a small Drude term.  At $n$ = 1, where eventually the charge reaches a uniform distribution throughout the film, a dramatic softening of the MIR band triggers a DE-driven IMT  around room temperature. As $T$ lowers, the MIR band continues to increase in intensity and to displace toward zero frequency, assuming a large polaron character, thus providing low-energy states to the Drude term. The polaronic character of the charges is confirmed by an effective mass, which at $n$=1 is $m^*/m_{band}\sim7$. 
The optical data, once extrapolated to $\omega = 0$, provide an alternative determination of the superlattice resistivity, which is not affected by paths perpendicular to the interfaces. Probably for this reason, they do not show the low-$T$ upturn exhibited by dc  measurements. As a whole, this optical study of the (SMO)n/(LMO)2n system shows that such heterostructure has its own properties, characterizing a novel electronic state which is profoundly different from those of doped bulk manganites except for the smallest interlayer spacing at $n=1$, where however the metallization at $T_{IMT}$ takes place in the unusual absence of disorder.

\section*{Acknowledgment}
The authors wish to thank A.J. Millis 
for fruitful discussions. SL acknowledges hospitality provided by the Condensed Matter Division, Max Planck Research Department for Structural Dynamics, Center for Free- 
Electron Laser Science, during the preparation of this 
manuscript. CA and DGS acknowledge support from the 
Cornell Center for Materials Research (CCMR) with funding 
from the Materials Research Science and Engineering Center 
program of the National Science Foundation (cooperative 
agreement DMR 0520404). 

\vspace{10mm}

This document is the unedited Author's version of a Submitted Work that was subsequently accepted for publication in Nano Letters, copyright \copyright  American Chemical Society after peer review. To access the final edited and published work see http://pubs.acs.org/doi/abs/10.1021/nl1022628.


\begin{thebibliography}{}

\bibitem{tokura99} Y. Tokura and Y Tomioka, J. Magn. Magn. Mater. {\bf 200}, 1 (1999). 
\bibitem{tokura06} Y. Tokura, Rep. Prog. Phys {\bf 69}, 797 (2006). 
\bibitem{zener51} C. Zener, Phys. Rev. {\bf 82}, 403 (1951).
\bibitem{millis95} A.J. Millis, P.B. Littlewood, B.I. Shraiman, Phys. Rev. Lett. {\bf 74}, 5144 (1995).
\bibitem{dagotto05} E. Dagotto, Science {\bf 309}, 257 (2005).
\bibitem{ohtomo02} A. Ohtomo, D.A. Muller, J.L. Grazul, and H.Y. Hwang, Nature {\bf 419}, 378 (2002).
\bibitem{okamoto04} S. Okamoto and A.J. Millis, Nature {\bf 428}, 630 (2004).
\bibitem{seo07} S.S.A. Seo, W.S. Choi, H.N. Lee, L. Yu, K.W. Kim, C. Bernhard, and T.W. Noh, Phys. Rev. Lett. {\bf 99}, 266801 (2007).
\bibitem{smadici07} S. Smadici {\it et al.}, Phys. Rev. Lett. {\bf 99}, 196404 (2007).
\bibitem{bhattacharya08} A. Bhattacharya {\it et al.}, Phys. Rev. Lett. {\bf 100}, 257203 (2008).
\bibitem{adamo09} C. Adamo {\it et al.}, Phys. Rev. B {\bf 79}, 045125 (2009).
\bibitem{aruta09} C. Aruta {\it et al.}, Phys. Rev. B {\bf 80}, 140405 (2009).
\bibitem{may09} S.J. May {\it et al.}, Nature Materials {\bf 8}, 892 (2009).
\bibitem{lin06} C. Lin, S. Okamoto, and A.J. Millis, Phys. Rev. B {\bf 73}, 041104(R) (2006).
\bibitem{lin08} C. Lin and A.J. Millis, Phys. Rev. B {\bf 78}, 184405 (2008).
\bibitem{dong08} S. Dong {\it et al.}, Phys. Rev. B {\bf 78}, 201102 (2008).
\bibitem{dressel} M. Dressel and G. Gr\"uner, in {\itshape Electrodynamics of Solids}, Cambridge University Press (2002). 
\bibitem{lupi09} S. Lupi {\it et al.}, Phys. Rev. Lett. {\bf 102}, 206409 (2009)); A. Perucchi et al., J. Phys. Cond. Matt. {\bf 21}, 323202 (2009), and references therein; S. Lupi {\it et al.}, Nat. Commun. 2010, 1:105. (doi: 10.1038/ ncomms1109). 
\bibitem{haghiri08} A.M. Haghiri-Gosnet {\it et al.}, Phys. Rev. B {\bf 68}, 115118 (2008).
\bibitem{dore97} P. Dore, G. De Marzi, and A. Paolone, Int. J. Infrared MM waves \textbf{18}, 125 (1997).
\bibitem{kuzmenko05} A.B. Kuzmenko, Rev. Sci. Instrum. {\bf 76}, 083108 (2005).
\bibitem{takenaka02} K. Takenaka {\it et al.}, Phys. Rev. B {\bf 65}, 184436 (2002).
\bibitem{calvani01} P. Calvani, {\it Optical Properties of Polarons}, Riv. Nuovo Cimento {\bf 24}, 1 (2001).
\bibitem{takenaka99} K. Takenaka {\it et al.}, Phys. Rev. B {\bf 60}, 13011 (1999).
\bibitem{mott90} N.F. Mott, {\it Metal-Insulator Transitions}, 2nd ed. (Taylor \& Francis, London, 1990).
\bibitem{okimoto97} Y. Okimoto {\it et al.}, Phys. Rev. B {\bf 55}, 4206 (1997).
\bibitem{nucara} A. Nucara {\it et al.}, unpublished.
\bibitem{mannella05} N. Mannella {\it et al.}, Nature {\bf 438}, 474 (2005).
\bibitem{vanmechelen08} J.L.M. van Mechelen {\it et al.}, Phys. Rev. Lett. {\bf 100}, 226403 (2008).
\bibitem{devreese09} J.T. Devreese and A.S. Alexandrov, Rep. Prog. Phys. {\bf 72}, 066501 (2009).


\end{thebibliography}
\end{document}